%% file: ms.tex
\documentclass[12pt]{emulateapj}
\begin{document}
\def\eg{{\it e.g.}}
\def\ie{{\it i.e.}}
\newcommand{\tnm}[1]{\tablenotemark{#1}}
\newbox\grsign
\setbox\grsign=\hbox{$>$}
\newdimen\grdimen
\grdimen=\ht\grsign
\newbox\simlessbox
\newbox\simgreatbox
\setbox\simgreatbox=\hbox{\raise.5ex\hbox{$>$}\llap
     {\lower.5ex\hbox{$\sim$}}}\ht1=\grdimen\dp1=0pt
\setbox\simlessbox=\hbox{\raise.5ex\hbox{$<$}\llap
     {\lower.5ex\hbox{$\sim$}}}\ht2=\grdimen\dp2=0pt
\def\simgreat{\mathrel{\copy\simgreatbox}}
\def\simless{\mathrel{\copy\simlessbox}}

\title{The Magellanic Bridge: The Nearest Purely Tidal Stellar Population}

\author{Jason Harris}
\affil{Steward Observatory}
\affil{933 North Cherry Ave., Tucson, AZ, 85721}
\email{jharris@as.arizona.edu}

\begin{abstract}
We report on observations of the stellar populations in twelve fields
spanning the region between the Magellanic Clouds, made with the
Mosaic-II camera on the 4-meter telescope at the Cerro-Tololo
Inter-American Observatory.  The two main goals of the observations
are to characterize the young stellar population (which presumably
formed {\it in situ} in the Bridge and therefore represents the
nearest stellar population formed from tidal debris), and to search
for an older stellar component (which would have been stripped from
either Cloud as stars, by the same tidal forces which formed the
gaseous Bridge).  We determine the star-formation history of the young
inter-Cloud population, which provides a constraint on the timing of
the gravitational interaction which formed the Bridge.  We do not
detect an older stellar population belonging to the Bridge in any of
our fields, implying that the material that was stripped from the
Clouds to form the Magellanic Bridge was very nearly a pure gas.
\end{abstract}


\section{Introduction}\label{sec:intro}
Interactions are known to be an important driver of galaxy evolution,
but a detailed understanding of their influence remains elusive.  The
Magellanic Clouds are a particularly compelling target for
investigating the effects of minor so-called ``harassment''
interactions, due to their proximity to the Milky Way, their close
association over at least the past several Gyr, and their abundant gas
reservoirs, which allow for ongoing star formation.  The strongest
evidence that their interaction has played an important role in
driving the evolution of the Clouds lies in the extra-tidal features of
the Magellanic Stream and Magellanic Bridge.  Unlike the Magellanic
Stream, which appears to be a pure-gas feature \citep{gr98}, there is
a known stellar population associated with the Magellanic Bridge, and
by measuring the ages, chemical abundances, and kinematics of these
stars, we can obtain strong constraints on the evolution of the
dynamical event which formed the Bridge, and study in detail how star
formation proceeds in the wake of such an event.

The Magellanic Bridge was first reported in \ion{H}{1} observations by
\cite{hkm63}, and a young stellar component (the ``inter-Cloud
population'') was discovered by \cite{ikd85}, who estimated the age of
the stars at about $10^8$~yr.  \cite{db98} provided the most
comprehensive study of the inter-Cloud population to date; they
observed five fields in the western Bridge, and found stars as young
as 10--25~Myr in both clusters and in a diffuse field component up to
$9^\circ$ from the SMC.  The young inter-Cloud population probably
formed {\it in situ} in the Bridge, in the wake of the Bridge-forming
event, making it the nearest example of a stellar population whose
formation was unambiguously triggered by a tidal interaction.  Yet
surprisingly, no detailed analysis of the star formation history of
these stars has been performed to date, and we do not even know the
full extent of the population, since no fields along the ridgeline of
the \ion{H}{1} gas have been observed east of the midpoint between the
Clouds.

Furthermore, no study to date has specifically searched for an older
component of the inter-Cloud population, which would represent a
population of stars that was stripped from the Clouds during the
Bridge-forming event.  Tidal forces during an interaction should
affect both gas and stars, so the inter-Cloud population
should have an old component, other things being equal.

The present study will address these open questions regarding the
inter-Cloud population in the Magellanic Bridge.  We present our
observations and data reduction in Section~\ref{sec:obsdata}.  In
Section~\ref{sec:youngstars}, we trace the eastward extent of the
young inter-Cloud population, and in Section~\ref{sec:oldstars}, we
search for tidally-stripped stars in our Bridge fields.  We briefly
examine the outer structure of the LMC using our four fields nearest
that galaxy in Section~\ref{sec:expdisk}.  Finally, we present the
star-formation history of the young inter-Cloud population in
Section~\ref{sec:sfh}, and summarize the results in
Section~\ref{sec:summary}.

\section{Observations and Data Reduction}\label{sec:obsdata}

\subsection{The Observations}

\begin{figure}[h]
\plotone{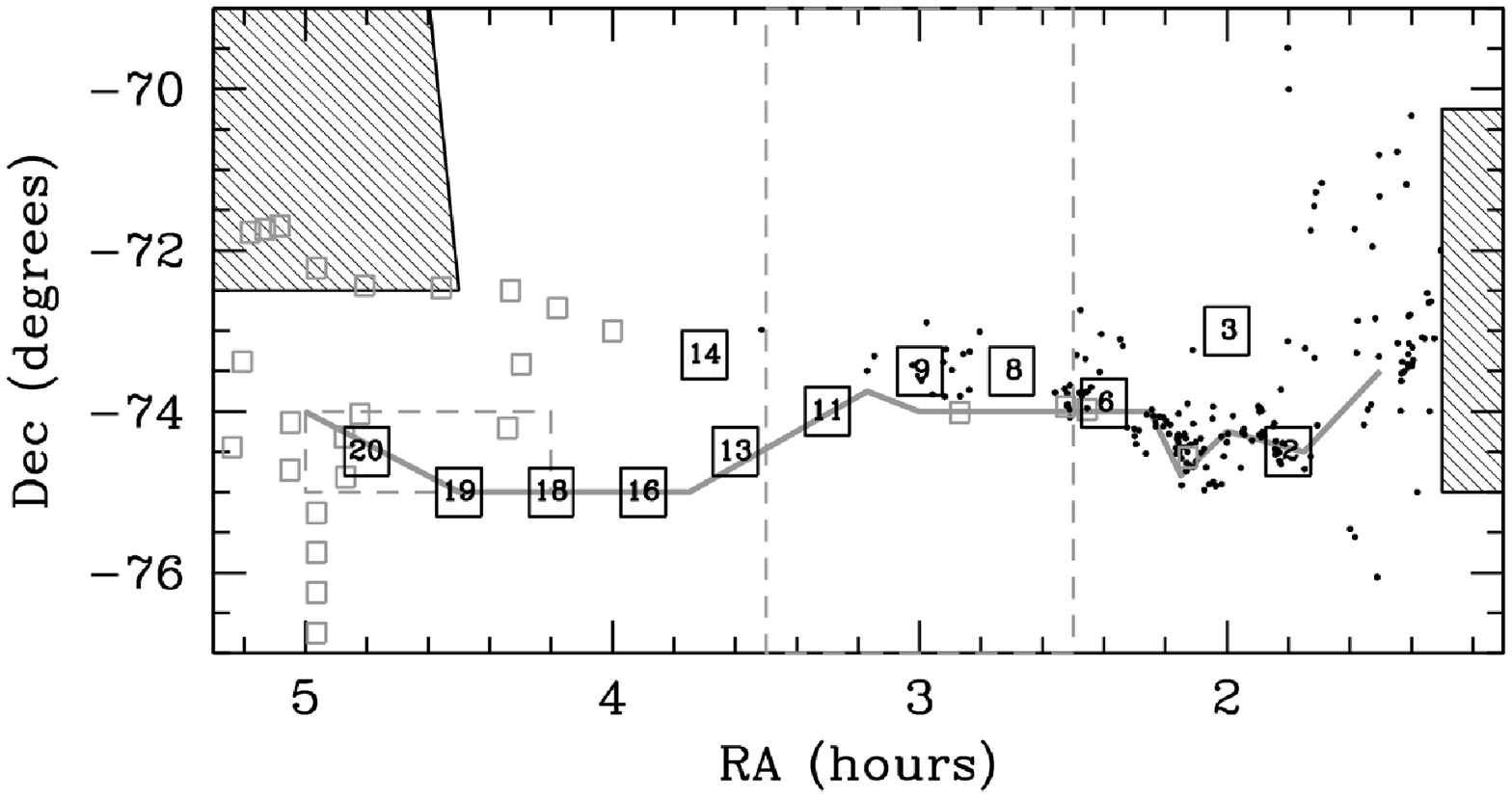}
\caption{A schematic view of the Magellanic Bridge region. The 
positions of the twelve observed CTIO-4m fields are shown as numbered 
boxes.  The inter-Cloud fields previously observed by \cite{db98} are 
shown as gray boxes.  Star clusters from \cite{bs95} are shown as 
points.  The gray line is an approximate trace of the HI ridgeline 
through the inter-Cloud region, from Figure~1 in \cite{put00}.  The 
hashed areas represent the portions of the regions covered by the 
Magellanic Clouds Photometric Survey (LMC at left, \citealp{zar04}; 
SMC at right, \citealp{zar02}). \label{fig:fields} }
\end{figure}

The data were obtained on the nights of January 4 and 5 2006 (UT), at
the Cerro-Tololo Inter-American Observatory (CTIO) 4-meter telescope.
We used the Mosaic-II camera, which images a
36$^\prime\times$36$^\prime$ field onto a 8k$\times$8k CCD detector
array, to obtain short and long exposures in Washington~$C$,
Harris~$R$, and Cousins~$I$ filters at twelve field positions spanning
the inter-Cloud region, and at one offset field at a similar Galactic
latitude, but to the west of the SMC (see Table~\ref{tab:exposures}
and Figure~\ref{fig:fields}).

The field positions were selected to uniformly sample the inter-Cloud
region, approximately following the ridgeline of the \ion{H}{1} gas
which forms the Magellanic Bridge \citep{put00}.  In addition, we
selected two fields to lie off the main ridgeline, but still in
regions of abundant \ion{H}{1} emission (fields mb03 and mb14).  Our
fields include the eastern half of the inter-Cloud region, where there
are very few known star clusters, and where \citeauthor{db98}'s fields
mostly lie far from the \ion{H}{1} ridgeline.

The exposure times listed in Table~\ref{tab:exposures} were chosen in
order to detect stars as faint as the ancient main-sequence turn-off
with S/N$>10$ in all three filters.  At each field position and for
each of the $C$, $R$, and $I$ filters, we obtained a pair of long
exposures for cosmic-ray rejection, and a short exposure to record the
photometry of brighter stars that are saturated in the long exposures.
Seeing was stable during both nights; the FWHM varied between
0.7$^{\prime\prime}$ and 1.0$^{\prime\prime}$.  We observed standard
star fields several times per night for photometric calibration (see
Table~\ref{tab:standards}), twilight flats were obtained during
evening and morning twilight of both nights, and bias frames were
obtained in the afternoon prior to both nights.

\subsection{Data Reduction}\label{sec:reduction}

The data reduction followed the procedure documented by the NOAO Deep
Wide Field Survey team \citep{jan03}, and utilized the {\tt mscred}
package in IRAF\footnote{IRAF is the Image Reduction and Analysis
  Facility, distributed by the National Optical Astronomy
  Observatories, which are operated by the Association of Universities
  for Research in Astronomy, Inc., under cooperative agreement with
  the National Science Foundation.}.  Before reducing the data, we
obtained an updated world coordinate system database for the CTIO
4-meter telescope from the CTIO website, dated from May 2004, and we
also obtained an updated crosstalk-correction parameter file.

We first processed the bias frames using {\tt ccdproc} and {\tt 
zerocombine}.  We then used {\tt ccdproc} to perform overscan 
correction and trimming, bias subtraction, bad pixel masking, 
amplifier merging and crosstalk correction on all data images.

We ran {\tt objmasks} on each twilight-sky exposure to mask out any
detected stars in the images, and then combined the exposures using
{\tt sflatcombine}.  This yielded normalized $C$, $R$, and $I$
twilight-sky flats for each observing night.  We then divided each
data frame by the appropriate normalized flat, using {\tt ccdproc}
with the {\tt sflatcor} parameter activated.  To further flatten the
data images, we also constructed night-sky flats from the collection
of data images, after running {\tt objmasks} on each image to mask out
detected objects.  We then ran {\tt ccdproc} once again to apply the
night-sky flats.

We empirically determined the world coordinate system (WCS) for each 
exposure using {\tt msccmatch}, which uses an astrometric database to 
identify stars in the image and determine the WCS from their known 
coordinates.  We chose to use the Guide Star Catalog, version 2 for 
the astrometric database, which we found yielded fit residuals that 
were smaller by about a factor of two compared to the USNO-A2 
catalog.  {\tt msccmatch} provides an interactive interface with 
which the fit can be improved by eliminating outlying mismatched 
points.  We were able to get the fit residuals in each image below 
0.25~arcsec.

Next, we used {\tt mscstack} to combine each pair of long exposures
into a single, averaged image.  We had initially activated the
cosmic-ray rejection feature of {\tt mscstack}, but we found that if
the seeing changed between the two exposures, the stellar flux in the
averaged image would be significantly clipped by the rejection
algorithm.  Instead, we stacked the images without rejection, and 
used {\tt craverage} to detect cosmic-ray hits in the combined image 
and replace the affected pixels with the average of the surrounding 
pixel values.  We also ran {\tt mscstack} on the short-exposure 
images, despite the fact that they did not have multiple exposures to 
combine, because {\tt mscstack} also performs a pixel-value 
replacement for the pixels in the bad pixel masks.

The final step in the reduction pipeline is to run {\tt mscpixarea},
which corrects each image for the variable pixel scale across the
field.  We then use {\tt mscsplit} to separate the eight CCD images
per field into separate FITS files.  We choose to analyze the CCD
images separately, in order to properly account for slight differences
between the CCDs.

\subsection{Instrumental Photometry}


We used the {\tt daophot} package in IRAF to perform stellar
photometry on our images, using the method of point-spread function
(PSF) fitting.  For each CCD image, we ran {\tt daofind} to identify
peaks, and {\tt phot} to obtain preliminary relative photometry of the
detected sources.

Next, we used {\tt pstselect} to select fifty bright, relatively
isolated sources to form a high-quality sample from which an
empirical PSF model will be built.  Candidate PSF stars are rejected 
if they contain saturated pixels, or if their centers are within ten 
pixels of any pixel flagged as bad by our data reduction procedure.

The PSF fitting is performed in a fully automated way, without user
intervention.  However, we do examine residual images after PSF
subtraction to ensure that the automatic model parameters are correct.
All available forms for the PSF are explored to find the best-fitting
model.  Next, {\tt group} and {\tt nstar} are run to identify and
photometer stars that are close neighbors of the PSF sample.  {\tt
  Substar} is used to subtract these close neighbors from the image.
We then repeat the PSF fitting, this time using the
neighbor-subtracted image to produce a more accurate model.

Once the second-pass PSF has been determined, we run {\tt allstar} to
perform iterative PSF photometry of all detected sources in the image.
Then we run {\tt daofind} on an image in which all known objects have
been subtracted using the PSF model, in order to find fainter stars in
the image.  Then {\tt allstar} is run on the subtracted image, using
this new list of fainter stars\footnote{Alternatively, we could have
performed the second-pass allstar on the original image using a
concatenated list of all detected objects.  We tested both methods on
an image from mb20, the most crowded field.  The differences in the
photometry are consistent with the uncertainties estimated by
allstar}.  The two allstar photometry lists are combined into a single
photometry table for the frame, and the IRAF task {\tt wcsctran} is
used to convert the stars' X,Y pixel coordinates to right ascension
and declination, using the world coordinate solution we determined
during the data reduction procedure.

The above photometry procedure results in some spurious detections due
to two artifacts: bleed columns, and scattered-light halos around
extremely bright stars.  The bleed columns are flagged as bad pixels
by the data reduction procedure, but this does not prevent {\tt
  daofind} from identifying sources along the bleed columns, nor does
it prevent {\tt allstar} from trying to photometer these false
sources.  We therefore remove objects from the photometry table that
are within 10 pixels of a flagged bad pixel.  Extremely bright stars
have large scattered-light halos in the images, resulting in circular
concentrations of false detections centered on these stars due to the
elevated flux levels.  To clean the photometry tables of sources
detected in the wings of extremely bright stars, we reject all sources
with anomalously high estimated sky values.  In the absence of these
extremely bright stars, the sky levels are quite stable, making the
sky levels an efficient way to identify spurious objects in the wings 
of bright objects.

The {\tt allstar} program provides very accurate relative photometry,
but because the PSF models are uncertain at large radii, the allstar
magnitudes are normalized to a relatively small aperture size of 4
pixels.  We therefore need to apply an aperture correction to convert 
the allstar magnitudes to true instrumental magnitudes that represent 
the total flux recorded from each star.  To determine the aperture 
correction, we select bright, isolated objects from our sample and 
perform concentric-aperture photometry using {\tt phot} with a series 
of aperture sizes, up to 17 pixels.  The per-star aperture correction 
is simply the difference between the star's allstar magnitude and its 
large-aperture magnitude: $(m_{als} - m_{ap17})$.  However, since 
individual objects suffer from measurement errors and contamination 
from neighboring objects, we need to statistically determine the 
characteristic aperture correction for the entire frame.  The
distribution of per-star aperture-correction values is typically a 
Gaussian with an asymmetric tail to negative values.  The Gaussian 
spread is due to measurement uncertainties, and the asymmetric tail 
is due to flux contamination, which is never completely mitigated by 
our selection of isolated objects.

An accurate determination of the aperture correction's value and
uncertainty requires that we attempt to isolate the underlying
Gaussian shape from the asymmetric skew caused by contaminating flux.
To do this, we first determine the approximate position of the 
distribution's peak, and then fit a Gaussian function to the points 
to the positive side of this peak value, thereby ignoring the 
negative half that may suffer from flux contamination.  We adopt the 
central value of the fitted Gaussian as the frame's aperture 
correction, and its width as the uncertainty in the frame's aperture 
correction.  The instrumental magnitude of each star is simply its 
allstar magnitude plus the frame's aperture correction; we also add 
the aperture correction uncertainty in quadrature to each star's 
photometric uncertainty.

\subsection{Standard Star Observations and Photometric Calibration}

The final step in our determination of the stellar photometry is to
place the instrumental magnitudes we have measured onto a standard
photometric system, using standard-star observations.  Standard star
fields were imaged several times on both of the observing nights.  One
standard field (SA~101) was observed at two separate visits on each
night in order to measure the effect of atmospheric extinction.

The standard star observations are presented in
Table~\ref{tab:standards}.  We selected well-known standard fields,
first measured by \cite{lan73}, expanded for wide-field CCD
instruments by \cite{ste00}, and calibrated for the Washington~$C$
filter by \cite{gei96}.  The standard fields were reduced using the
pipeline procedure described in Section~\ref{sec:reduction}.  We then
identified sources in each standard field with {\tt daofind}, and
performed concentric-aperture photometry on all sources with {\tt
  phot} in IRAF's {\tt daophot} package.

We use the measured aperture photometry of the observed standard
stars, together with their total photometry as published by
\cite{ste00} and \cite{gei96}, to solve the following photometric
calibration equation for each filter and each CCD in the Mosaic-II
array, and independently for the two nights of observing:

$$M = m + A + B*(R-I) - C*X$$

where $M$ is the published total magnitude, $m$ is the observed
instrumental aperture magnitude, $A$ is the photometric zeropoint, $B$
is the color term, $R-I$ is the star's true color, $C$ is the
atmospheric extinction term, and $X$ is the airmass.  Once we have
determined $A$, $B$ and $C$ for each CCD and filter, we will be able
to convert the observed photometry of any star to its total
photometry, given its observed color and the airmass at which it was
observed.

We first determine $C$, the atmospheric extinction term, by examining
the photometry from SA~101, the standard field that was observed at
different airmasses during each night.  The $C$ parameter is
independent of the color term and zeropoint, so we simply need to fit
a linear regression through the observed stars' magnitudes as a
function of the observed airmass.  The slope of the linear regression
is $C$, the atmospheric extinction term.  Note that we need not
restrict ourselves to the actual standard stars for this step; since
we only need the relative photometry to determine the extinction term,
all of the stars observed in field SA~101 can be employed.

Having determined $C$, we proceed to simultaneously determine the
zeropoint and color term.  The observed magnitudes of all standard
stars are first corrected for atmospheric extinction ($m_x = m -
C*X$); we then construct the quantity $M-m_x$, the difference between
the the published total magnitude of the star and its
extinction-corrected instrumental magnitude, and fit a linear
regression through the $M-m_x$ values as a function of the published
$R-I$ colors.  The slope of this regression is the color term $B$, and
its zeropoint is $A$, the photometric zeropoint correction.
Table~\ref{tab:photcalib} presents the photometric calibration
parameters for each CCD and filter, and for each of the two observing
nights.

Note that in Table~\ref{tab:photcalib}, the parameters for the
Washington~$C$ filter are the same in all eight CCDs.  The reason for
this is that there are too few standard stars calibrated for the $C$
filter to support an independent determination for each CCD (see
Table~\ref{tab:standards}), so we were forced to determine an average $C$
calibration for the entire mosaic.

As noted in Table~\ref{tab:photcalib}, there were too few $R$ and $I$ 
standards present in some of the CCDs to support an independent 
determination of their color terms and zeropoints.  For these cases, 
we perform a bootstrap estimate of the parameter values from those 
published at the CTIO 
website\footnote{http://www.ctio.noao.edu/mosaic/ZeroPoints.html}.  
We determine the mean offset between our determined values of the
photometric calibration parameters, and those published by CTIO, for
the CCDs that we were able to analyze.  We then apply this mean offset
to the published values of the remaining CCDs, as an estimate of what
we would have measured if we had observed enough standard stars in
those CCDs.  For the uncertainty in these bootstrapped parameters, we
simply adopt the standard deviation of the mean offset between the
observed and published values.

\begin{figure}[h]
\plotone{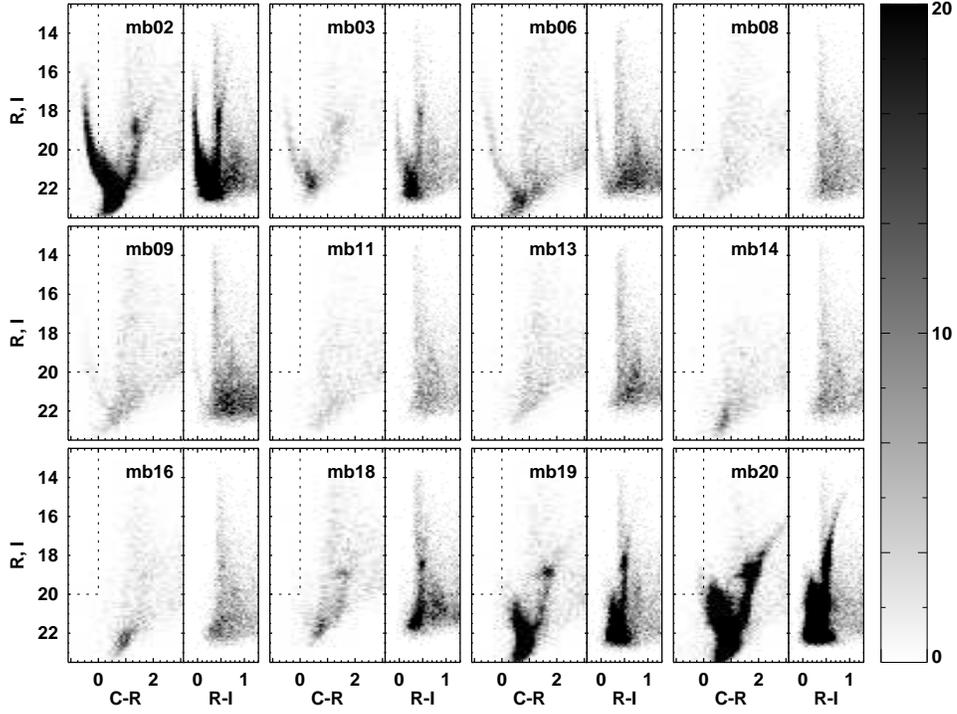}
\caption{Color-magnitude Hess diagrams of each of the twelve fields
observed in the Magellanic Bridge.  Each pair of panels consists of
the $C-R$,$R$ CMD (left) and the $R-I$,$I$ CMD (right).  The fields
are numbered in order of increasing right ascension (see
Figure~\ref{fig:fields}).  Dashed lines indicate the region of the
$C-R$ CMDs where main sequence stars younger than 1~Gyr are
expected. The pixel values indicate the number of stars present in
each pixel, ranging from zero (white) to 20 (black), as indicated by
the scale bar on the right. \label{fig:cmds} }
\end{figure}

We simultaneously perform a positional match of the sources in the
$C$, $R$ and $I$ photometry lists, and apply the above photometric
calibration to produce catalogs with total $CRI$ photometry.  We then 
match objects between the $CRI$ catalogs from the short and long 
exposures of each field.  For the positional matching, we use
a maximum match radius of 0.5$^{\prime\prime}$; to be retained in the
catalog, an object must be detected in the $R$ band, with a matching
detection in either $C$ or $I$.  For objects which are matched between
the short and long exposures, we adopt the weighted mean photometry in
the final catalog; objects present only in the short or long catalog 
are included as well.  The final calibrated composite photometry 
catalogs for the twelve observed fields in the Magellanic Bridge 
(plus the observed offset field) are presented in 
Table~\ref{tab:catalog}, and rendered as pairs of Hess diagrams in 
Figure~\ref{fig:cmds}.  A Hess diagram is a pixelized color-magnitude 
diagram (CMD) in which each pixel value is proportional to the number 
of stars in the region covered by that pixel.  The CMD of the offset 
field is presented separately in Figure~\ref{fig:offset-cmd}.

\subsection{Statistical Subtraction of Foreground/Background Contamination}\label{sec:statsub}

\begin{figure}[h]
\plotone{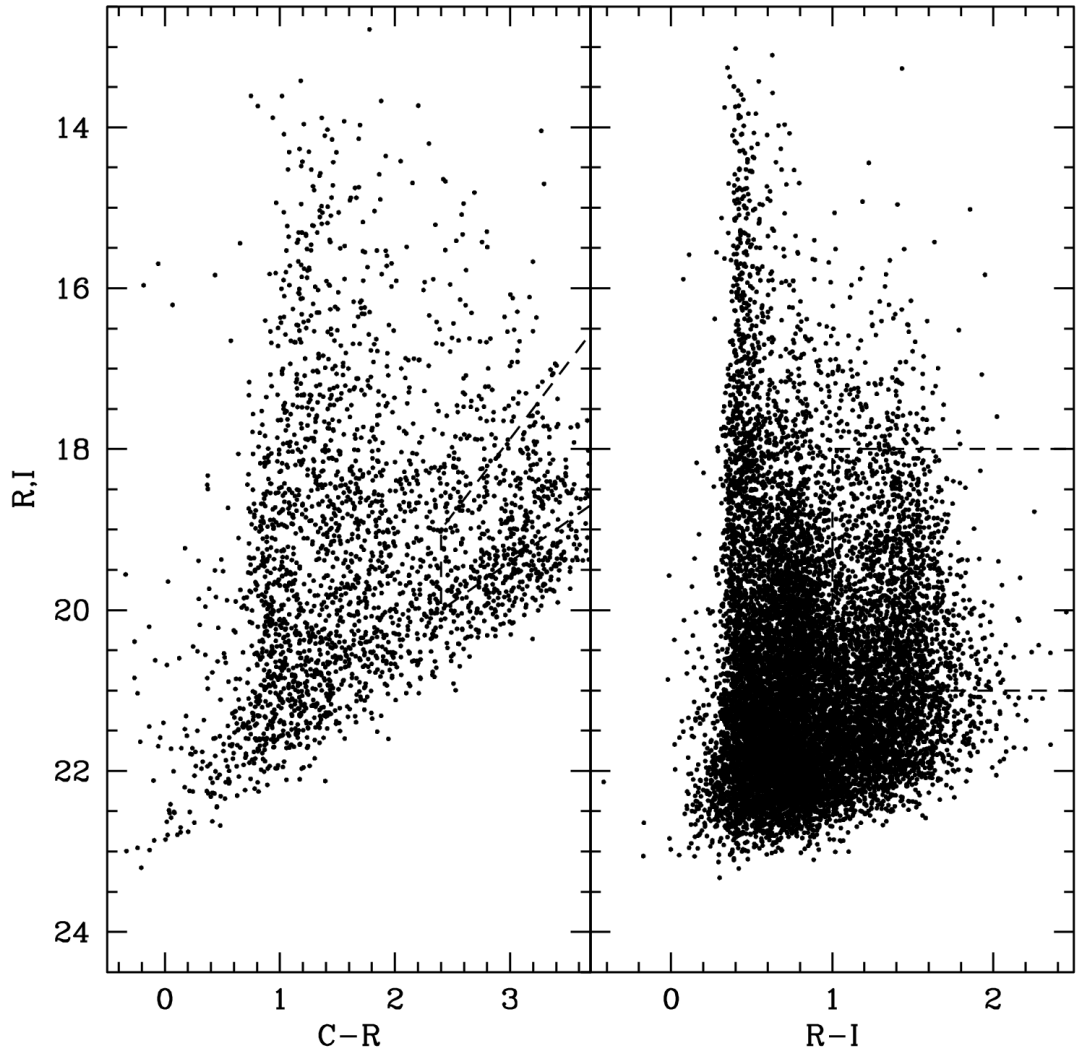}
\caption{The CMD pair for the offset field, which has a similar
Galactic latitude to the Bridge fields, but lies to the west of the
SMC.  It therefore should represent a pure foreground Galactic and
background extragalactic population.  As in Figure~\ref{fig:cmds}, the
$C-R$,$R$ CMD is shown at left, and the $R-I$,$I$ CMD is shown at
right.  The normalization regions, which are expected to contain only
contaminant objects, are outlined with dashed lines.
\label{fig:offset-cmd} }
\end{figure}

It is clear from comparing Figures~\ref{fig:cmds} and 
\ref{fig:offset-cmd} that the stellar populations in many of the 
fields are dominated by foreground Galactic (and background 
extragalactic) contamination.  In order to study the underlying 
inter-Cloud populations, we need to first perform a statistical 
subtraction of the contaminant foreground/background population.  In 
doing so, we will assume that the population observed in the offset 
field is representative of the contaminant population in each Bridge 
field (a reasonable assumption, given the similar Galactic latitude 
of the offset field).

\begin{figure}[h]
\plotone{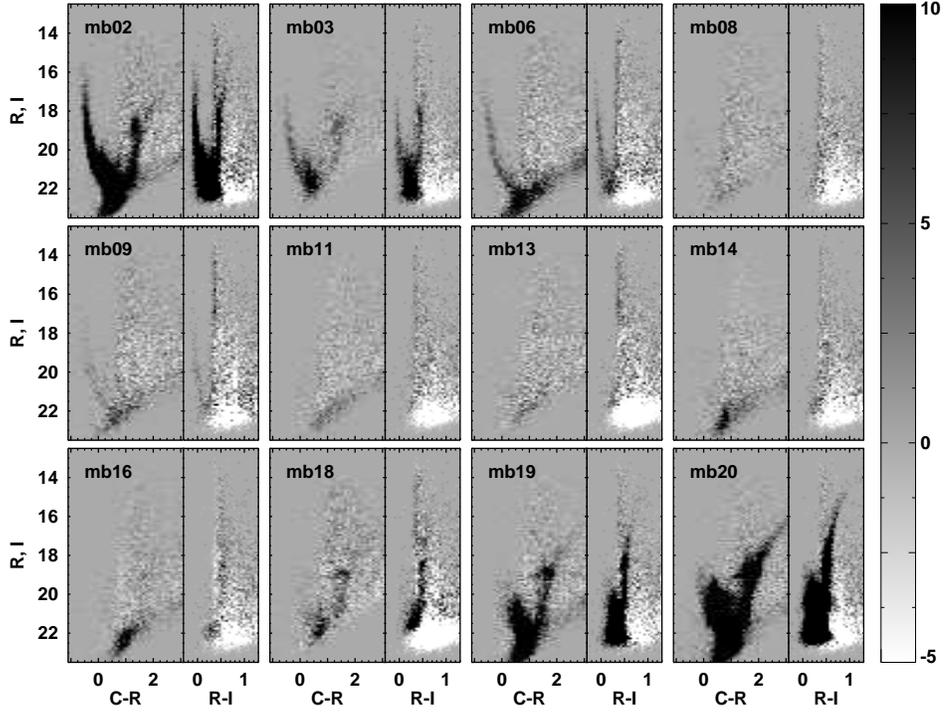}
\caption{Color-magnitude Hess diagrams of the twelve observed Bridge
fields, after statistically subtracting a contaminant component
derived from the offset field population (see
Figure~\ref{fig:offset-cmd}).  Darker pixels indicate a local excess
of stars compared to the contaminant population, whereas lighter
pixels indicate a local deficit of stars.  In a perfect subtraction,
the pixels lighter than the medium gray of zero difference would be
consistent with the noise of the subtraction.  Many of the fields show
large deficits of stars at the faint end of the $R-I$ CMD; we believe
this is due to a mismatch in the faint limit between these target
fields and the offset field.  The pixel values indicate the difference
in star counts $(N_{field} - N_{offset})$ for each pixel, ranging from
-5 (white) to +10 (black), as indicated by the scale bar on the right.
\label{fig:clean-cmds} }
\end{figure}

We proceed by first determining a scaling factor for normalizing the
number of objects in the offset field to the number of contaminant
objects in each Bridge field.  This is necessary to account for
variations in the effective area covered by each field (which arise
from masking out regions contaminated by bad pixels, very bright stars
and bleed trails).  The normalization factor is simply the ratio of
object counts in the target and offset fields, for a selected
subregion of each CMD that is expected to contain only contaminant
objects.  For the $C-R$ CMD, the normalization region is defined by
the criteria $C-R > 2.4$, $R < 22.4 - (C-R)$, and $R > 23.8 -
2*(C-R)$; for the $R-I$ CMD the criteria are $R-I > 1.0$ and $18 < I <
20$.  The normalization regions are outlined with dashed lines in
Figure~\ref{fig:offset-cmd}.  In the $C-R$ CMD, the objects in the
normalization region constitute 12\% of the total number of objects,
while in the $R-I$ CMD, the fraction is 14\%.  The normalization
factor computed for each target field ranges between 0.65 and 0.95.
We multiply the offset field's Hess diagrams by these scaling factors,
and subtract them from the target field's Hess diagrams.  The
resultant statistically-cleaned Hess diagrams for each field are shown
in Figure~\ref{fig:clean-cmds}.

The statistical subtraction was generally successful in removing a
component from each field's CMD that is consistent with the
contaminant population in the offset field.  However, there are some
artifacts present that bear explanation.  Specifically, the faint end
of many of the $R-I$ CMDs appear to show an oversubtracted contaminant
population.  This is simply due to the fact that the offset field's
$R-I$ CMD has a fainter detection limit than that in most of the
target fields.

\section{Analysis}\label{sec:analysis}

\subsection{Extent of the Young Inter-Cloud Population}\label{sec:youngstars}

Young (age $<1$~Gyr) stars provide an unambiguous tracer of the
inter-Cloud population, because no foreground or background
contaminants are expected to share the bright, blue region of the CMD
with these stars.  From previous work on the young inter-Cloud
population by \cite{db98}, we expected to observe a population of
young stars near the SMC, coincident with the young cluster population
cataloged by \cite{bs95}.  However, we did not know how far the young
population would extend toward the LMC along the \ion{H}{1} ridgeline.
We isolate stellar populations younger than 1~Gyr in the $C-R$ CMD, by
selecting those stars with $R<20$~mag and $C-R<0$~mag (see dashed
lines in Figure~\ref{fig:cmds}).  Stars matching these criteria are
absent in all of our fields east of mb09, which corresponds roughly to
the eastern extent of the \citeauthor{bs95} clusters.  Interestingly,
field mb09 is also near the position along the \ion{H}{1} Bridge where
the gas surface density drops to the critical threshold for star
formation of 3--4~$M_\odot pc^{-2}$ \citep{ken89}, which corresponds
to the $5\times10^{21} cm^{-2}$ seen throughout the eastern Bridge in
Figure~4a of \cite{bru05}.  West of field mb09, the gas density is
sustained at a level three times higher, and this is where star
formation has been active in the Bridge.  It would seem that the same
star formation threshold observed for disk galaxies holds for this
tidal debris environment as well.

\subsection{Searching for Tidally-Stripped Stars in the Inter-Cloud Region}\label{sec:oldstars}

It is perhaps not surprising that the young inter-Cloud population is
confined to those regions where the gas density is relatively high, if
we accept the hypothesis that these stars formed {\it in situ},
following the formation of the gaseous Bridge by a recent
gravitational encounter between the Clouds.  However, the tidal forces
that presumably formed the Bridge should have stripped stars and gas
with equal efficiency, so we expect to observe a population of such
tidally-stripped stars in the inter-Cloud region.  \cite{yn03}
conducted detailed numerical modeling of the stars and gas in the SMC,
as it orbits both the LMC and Milky Way, in an attempt to reproduce
the broad physical parameters of the Magellanic system.  In their
best-fitting model, there is an abundant population of stars in the
inter-Cloud region which formed in the SMC, and had been ejected into
the inter-Cloud region by a tidal interaction with the LMC.  The
tidally-stripped stars should have a similar age distribution to the
stars in the galaxy from which they were stripped (at least for ages
prior to the Bridge-forming event when their histories diverged).
Since the stellar populations in both Magellanic Clouds exhibit a
prominent red giant branch and a ``red clump'' horizontal branch,
these bright features serve as ideal tracers of a putative stellar
population that had been stripped from either of the Clouds during the
Bridge-forming event.

While some of the fields in Figure~\ref{fig:clean-cmds} do show red
giant branch and red clump features, these older populations appear to
be confined to the fields nearest the SMC (fields mb02 and mb03) or
the LMC (fields mb16--mb20).  Furthermore, the surface density of
these tracer populations increases sharply as the galactocentric
separation of the field decreases, consistent with populations that
are bound to the LMC and SMC.  In Section~\ref{sec:expdisk}, we will
demonstrate that the red giant populations in fields mb16--mb20 are
consistent with a plausible exponential disk distribution centered on
the LMC.  For now, we simply conclude that the red giant populations
in these six fields near the SMC and LMC are very likely composed of
stars bound to each respective galaxy, and are not indicative of a
tidally-stripped stellar population in the Magellanic Bridge.

\begin{figure}[h]
\plotone{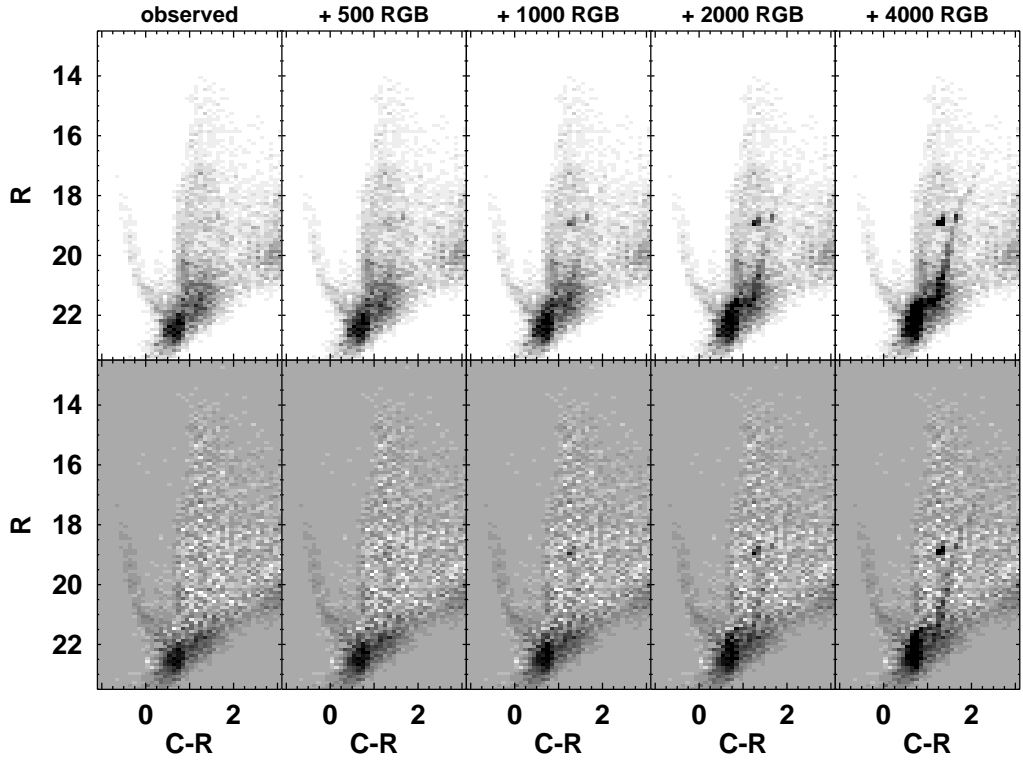}
\caption{The top left panel shows the $C-R$ Hess diagram for a 
composite population formed by combining the ``true Bridge'' fields 
(mb06--mb14).  We see no trace of a red giant branch feature which 
would indicate an older stellar population in these fields.  We 
determine the upper limit on the number of red giants which could 
remain undetected in this diagram by adding a synthetic red giant 
branch composed of increasing numbers of stars.  From left to right, 
starting with the second panel we show the same Hess diagram after 
having added 500, 1000, 2000, and 4000 synthetic red giant stars.  
The bottom row of panels show the same Hess diagrams, after having 
statistically subtracted a contaminant population derived from the 
offset field. 
\label{fig:fakergb} }
\end{figure}

The Hess diagrams of the remaining six fields (mb06--mb14) show no red
features that can be associated with an old inter-Cloud population.
However, the strength of this non-detection is limited by the presence
of the contaminant population.  To enhance our sensitivity to a
potentially sparse old stellar population, we construct a composite
pair of Hess diagrams from these six ``true Bridge'' fields, and
perform a new statistical contaminant subtraction on the composite
population (see left panels of Figure~\ref{fig:fakergb}).  Even in
this composite Hess diagram which covers more than two square degrees of
the inter-Cloud region, there is no detectable trace of an underlying
red giant branch or red clump feature.

We can place an upper limit on the surface density of red giant branch
stars in these six ``true Bridge'' by adding an artificial old stellar
population at the distance of the Magellanic system ($m-M=18.7$~mag,
intermediate between the two Clouds) to the composite inter-Cloud
population.  The artificial old stellar population is drawn from a
theoretical isochrone \citep{gir02} with $Z=0.002$ and
$log(age)=10.0$, to which we add photometric errors consistent with
the data.  We modulate the number of artificial stars added until a
red giant branch is marginally detectable (Figure~\ref{fig:fakergb}).
We conclude from this exercise that there are fewer than 1000 red
giant branch stars at the distance of the Bridge and brighter than
$R=23$~mag in the observed composite population.  By applying a
stellar mass function \citep{kro01}, we can convert the upper limit on
the number of observed red giants to an upper limit on the total
stellar mass present in a putative old stellar population.  However,
the conversion factor depends on the assumed age of the stars, because
the fraction of the total stellar population that is brighter than
$R=23$~mag varies with age.  For a 10~Gyr population, the upper mass
limit is 14800~$M_\odot$, and for a 2.5~Gyr population, the upper mass
limit is 5300~$M_\odot$.  Thus, the stellar surface mass density in
these six ``true Bridge'' fields is $\le0.009\ M_\odot\ pc^{-2}$; this
is more than 400 times smaller than the average surface mass density
of \ion{H}{1} in the Magellanic Bridge \citep[4~$M_\odot\ pc^{-2}$,
  converted from the characteristic column density in the
  Bridge,\ ][]{bru05}.  There does not appear to be any trace of a
tidally-stripped stellar population in the Magellanic Bridge, at least
in these six fields along the \ion{H}{1} ridgeline.

One potential caveat in this analysis is that we have assumed that the
putative tidally stripped stellar population would be spatially
coincident with the gaseous Bridge.  This need not be the case; if
ram-pressure from the Milky Way halo has played a significant role in
the evolution of the Magellanic system \citep{mas05}, then it is
possible that the gaseous Bridge is now displaced from the region
occupied by tidally-stripped stars between the Clouds.  We investigate
this possibility using data from the 2-Micron All-Sky Survey
\citep[2MASS, ][]{skr06}.

Using the Gator web-based database query
service\footnote{http://irsa.ipac.caltech.edu/applications/Gator} at
the NASA/IPAC Infrared Science Archive, we obtained near-infrared
$JHK$ photometry from the 2MASS All-Sky Point Source Catalog, in two
regions (shown as dashed boxes in Figure~\ref{fig:fields}).  The first
2MASS region (the ``full-bridge region'') was selected to cover all
plausible locations where a tidally-stripped inter-Cloud population
might exist.  We selected a range in right ascension between 2.5$^h$
and 3.5$^h$, because these limits are bracketed by fields mb06 and
mb14, which define the edges of the ``pure bridge'' section of our
sample, uncontaminated by LMC or SMC stars.  We selected a very large
range in declination, from $-77^\circ$ to $-69^\circ$, to cover all
plausible trajectories of a putative tidally-stripped stellar
population.  The second 2MASS region (the ``SW-LMC region'') was
selected as a comparison field that is known to contain an old stellar
population at the distance of the Magellanic system.  This region
spans 4.2$^h$ to 5$^h$ in right ascension, and $-75^\circ$ to
$-74^\circ$ in declination.  It is coincident with our fields mb18,
mb19 and mb20, in which we have observed an old stellar population
associated with the LMC (Section~\ref{sec:oldstars}).  While the
full-bridge region covers a solid angle ten times larger than that of
the SW-LMC region (35 square degrees and 3.2 square degrees,
respectively), the 2MASS catalog contains about the same number of
stars in both regions (94000 stars in the full-bridge region, and
92000 stars in the SW-LMC region), due to the larger stellar surface
density of the SW-LMC region.

\begin{figure}[h]
\plotone{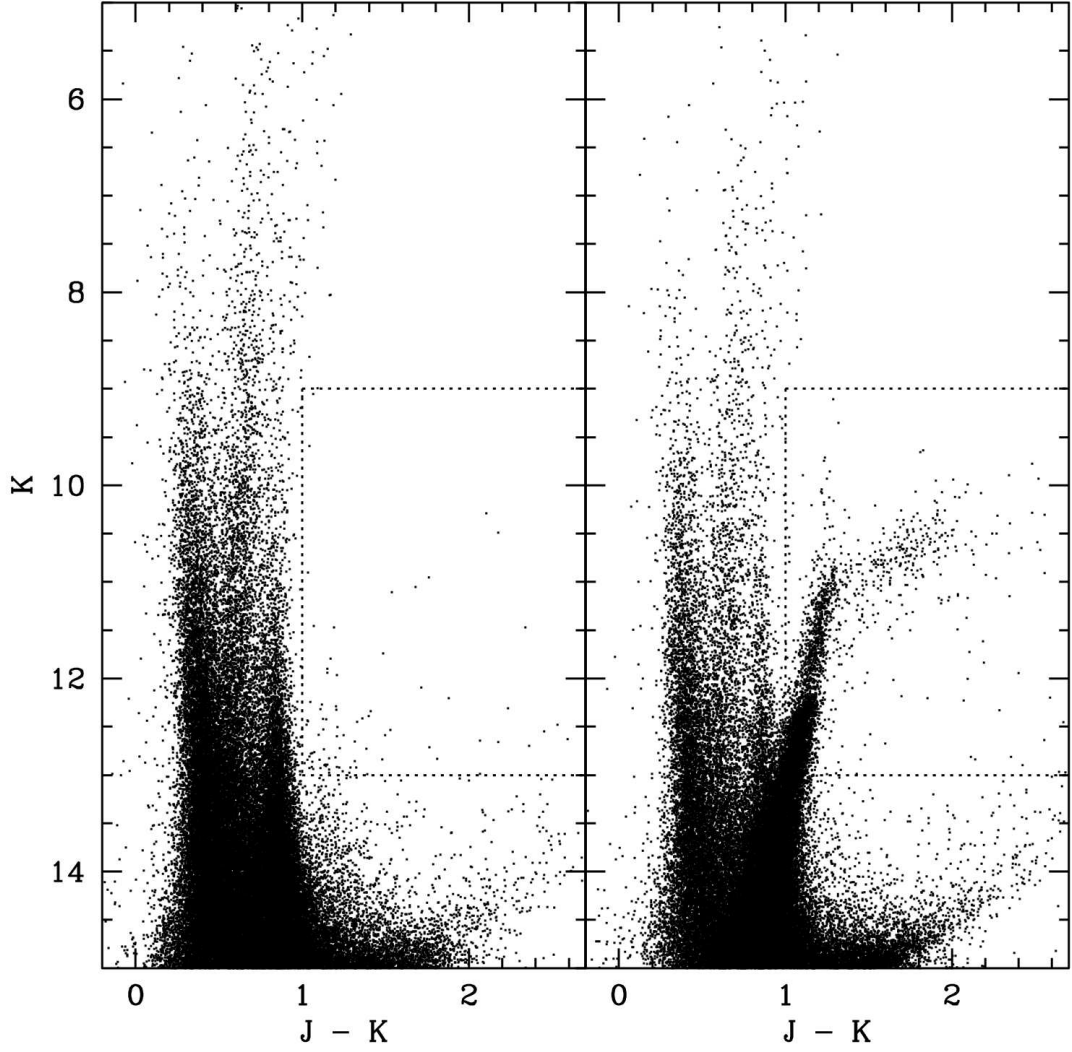}
\caption{Near-infrared $J-K$ color-magnitude diagrams for two regions
between the Magellanic Clouds.  At left is the ``full-bridge region'',
which spans 35 square degrees and is designed to cover the full region
between the Clouds that is not contaminated by stars bound to either
galaxy.  At right is the ``SW-LMC region'', which covers 3.2 square
degrees and is coincident with our fields mb18, mb19, and mb20, in
which we have observed an old stellar population associated with the
LMC.  The dotted lines in each panel indicate the region where the
bright red giant and asymptotic giant branches are expected at the
distance of the Magellanic system.  While an abundant population of
these stars is present in the SW-LMC region, the much larger
full-bridge region contains hardly any such stars.
\label{fig:2mass-cmds} }
\end{figure}

The 2MASS $J-K$ CMDs for these two regions are shown in
Figure~\ref{fig:2mass-cmds}.  In the SW-LMC region, there is an
abundant population of red objects consisting of old stars associated
with the the LMC.  Following \cite{nw00}, we identify the various
subpopulations of these red objects.  The bulk of the population
extends in a narrow diagonal sequence from $J-K$=1~mag,$K$=14~mag to
$J-K$=1.25~mag,$K$=11~mag.  Along this sequence, there is a sharp drop
in the density of stars around $K$=12.3~mag; this is the tip of the
red giant branch.  The stars in this sequence brighter than
$K$=12.3~mag are oxygen-rich asymptotic giants, while the stars which
extend redward of $J-K$=1.25~mag are carbon-rich asymptotic giants.
In the full-bridge region's $J-K$ CMD, there is a small number of
stars whose photometry is consistent with these features (notably the
$\sim6$ red objects around $K=11$~mag which may be Carbon stars at the
Magellanic distance), but considering the very large solid angle
covered by the full-bridge region, we do not regard these objects as a
significant detection of an old inter-Cloud population.  We can place
an upper limit on the number of red giants at the distance of the
Magellanic system that can remain undetected in the 2MASS CMD, using
the same synthetic population analysis described above for our optical
CMDs.  A red giant population containing 150 stars brighter than
$K=14$~mag is easily detectable when added to the 2MASS CMD, which
implies that any old inter-Cloud stellar population that may be
present has a total stellar mass no greater than $2\times10^6\
M_\odot$.  This is 1\% of the total \ion{H}{1} mass in the Magellanic
Bridge \citep{bru05}.  However, the area covered by our full-Bridge
region is about three times smaller than the area used to define the
Bridge by \citeauthor{bru05}, so the true limit from the 2MASS data is
closer to 3\%.  Thus, even accounting for the possibility that a
putative tidally-stripped stellar population may be displaced from the
gaseous Magellanic Bridge, we can still conclude that the Bridge
material was more than 97\% gas when the Bridge was formed.

\subsection{The Outer Disk of the Large Magellanic Cloud}\label{sec:expdisk}

\begin{figure}[h]
\plotone{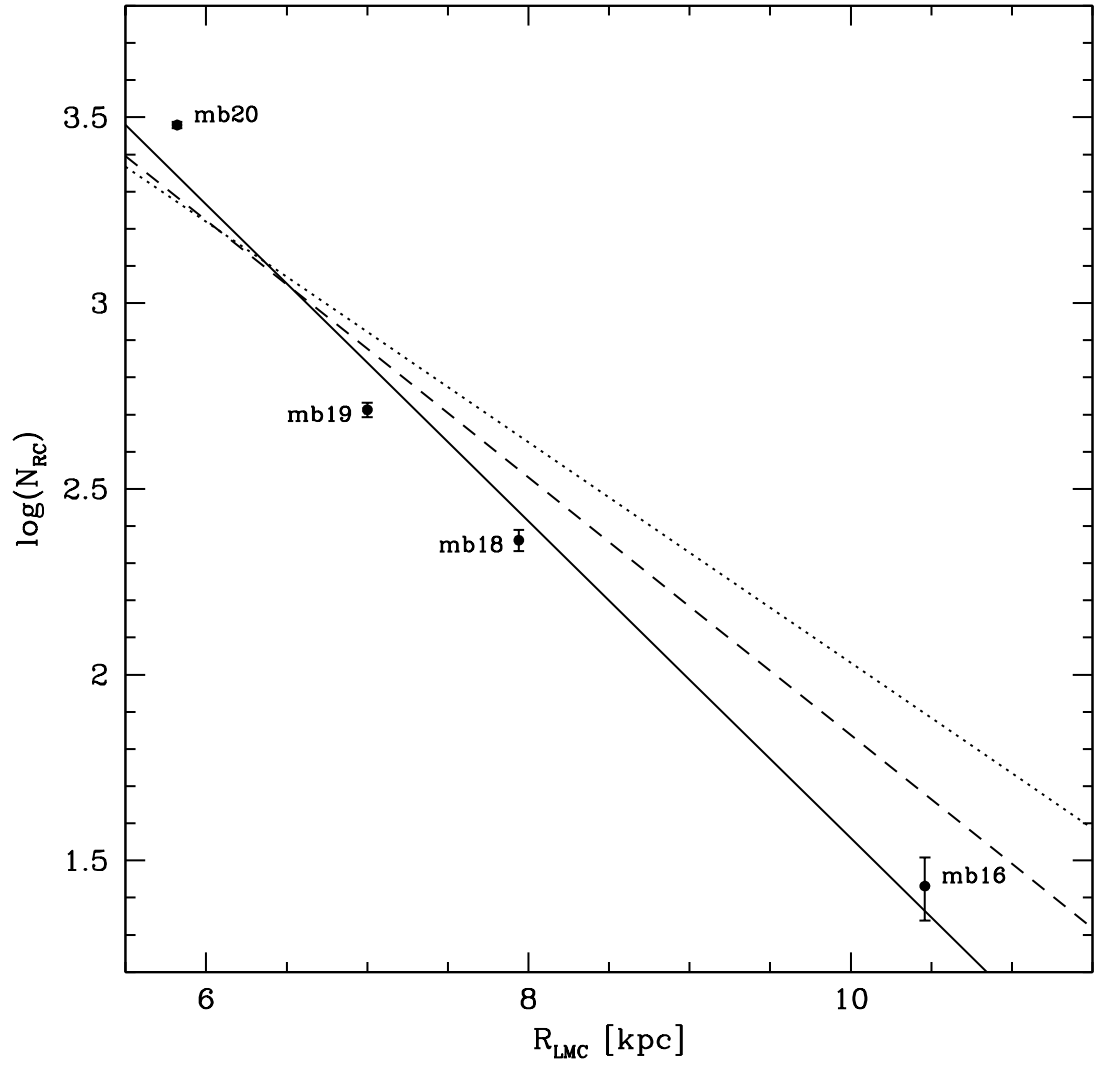}
\caption{The number of red clump stars present in fields mb16--mb20
vs. the physical distance of each field from the center of the LMC.  
In determining these distances, we have assumed that the red clump 
stars are in the disk of the LMC, whose geometry is known.  The 
best-fit exponential-disk model for these data is shown as the 
solid line, and it is similar to previous determinations of the 
LMC's exponential disk by \citep[][; dashed line]{har78} and 
\citep[][; dotted line]{gal04}.
\label{fig:expdisk} }
\end{figure}

In fields mb16--mb20 we observe old stellar populations that we
conclude are bound members of the LMC, based on the sharp increase in
their surface density with decreasing angular separation from the LMC.
\cite{gal04} and others have found that the LMC's stellar radial
profile follows an exponential disk to projected radii beyond 7~kpc,
with no sign of a break which might indicate the onset of a kinematic
halo.  Fields mb16--mb20 have projected separations from the LMC of
between 5~kpc and 8.5~kpc; however, when the orientation of the LMC
disk \citep{vdm01} is taken into account, the in-disk galactocentric
distances of these fields are between 6~kpc and 10.5~kpc.  We use the
number of stars in the red clump feature as a proxy for the stellar
surface density, and plot the surface density profile in
Figure~\ref{fig:expdisk}.  The solid curve represents the best-fit
exponential-disk model, with a scale length of $\alpha=0.98$~kpc, and
the dotted and dashed curves are exponential disk models fit by
previous authors, as noted in the figure caption.  The surface density
profile of the red clump stars in fields mb16--mb20 are generally
consistent with previous measurements of the LMC's outer exponential
disk, but the fact that the profile is somewhat steeper in this
southwestern quadrant is interesting.  A comprehensive survey of the
stellar populations in the outer LMC is currently underway; we will
therefore postpone further discussion of the LMC's structure until
this survey is completed, when more definitive conclusions can be
made.

\subsection{Characterizing the Purely Tidal Stellar Population in the Inter-Cloud Region}\label{sec:sfh}

We have determined that the stars in the inter-Cloud region appear to
be exclusively composed of a stellar population that formed {\it in
  situ}, in the wake of the Bridge-forming event (modulo some
contribution from stars still bound to the LMC and to the SMC, in the
observed Bridge fields nearest those galaxies).  This isolation of a
tidally-triggered stellar population provides an important opportunity
to examine the nature and evolution of star formation processes in
tidal debris.  We measure the age distribution of the inter-Cloud
population to determine when the star formation occurred, and how long
it lasted.  Since these stars presumably formed in the wake of the
Bridge-forming event, these measurements provide an important
constraint on the timing of that event.  We will also look for spatial
structure in the age distribution, which may provide insights into how
star formation proceeds when triggered by a gravitational interaction.

Previous studies of the inter-Cloud population have estimated the age
of the youngest stars present, through simple isochrone fitting
\citep[e.g.,\ ][]{db98}.  Here we will perform a more detailed
analysis, using the StarFISH star formation history fitter
\citep{hz01}.  This analysis is motivated by the clear presence of
composite stellar populations in some of our fields, and by our goal
to constrain the duration of star-formation activity in the Bridge.

StarFISH constructs a library of synthetic CMDs, each of which
represents a model of what the photometric observations would yield,
if the observed stellar population had a single age and a single
metallicity.  The model photometry is derived from theoretical
isochrones; in this case we chose the latest Padua isochrones
\citep{gir02}.  In order to accurately predict the observed
photometric distribution in the CMDs, the models include a distance
modulus, a distribution of extinction values, and a detailed model of
the photometric errors.  The distance modulus was simply chosen to be
that of the SMC, 18.9~mag, because the young stellar populations in
the Bridge are near the SMC on the sky.  The distribution of
extinction values is drawn from regions near the eastern edge of the
MCPS SMC extinction map \citep{zar02}.  For the photometric errors, we
employ an analytic model that reproduces the error statistics in the
observed fields.  While we usually advocate for an empirical model
based on artificial stars tests, these tests are only strictly
necessary when the data images are crowded.  In the present case, even
in our field with the highest stellar surface density (mb20), we have
detected roughly 109000 stars in $8192^2$ pixels, corresponding to a
mean separation between objects of almost 14~pixels.

The StarFISH model library provides synthetic CMDs for the range of
ages and metallicities thought to be present in the observed
population; in the present case we constructed synthetic CMDs for 16
age bins spanning ages 10~Myr to 12~Gyr, spaced uniformly in
$log(age)$, and for three metallicity bins, $Z=0.001$, $Z=0.002$, and
$Z=0.004$.  The best-fit SFH is found by determining the combination
of amplitudes modulating these synthetic CMDs which produces a
composite model CMD that most closely matches the observed CMD.  To
take advantage of the full $CRI$ photometric data set in determining
the SFH, the fit is actually performed on the CMD pair: $C-R$ vs. $R$
and $R-I$ vs. $I$.

Because the contaminant population dominates many of our observed
fields, it is important to account for contaminants in the SFH fit.
We could have run StarFISH on the statistically-cleaned data set, but
we instead chose to use the observed data set, and simply include the
contaminant offset-field population as an additional amplitude in the
model, in addition to the normal set of synthetic CMDs.  The code will
then select a multiplicative amplitude factor that optimally accounts
for the contaminant population, just as it does for each of the
synthetic CMDs.

\begin{figure}[h]
\plotone{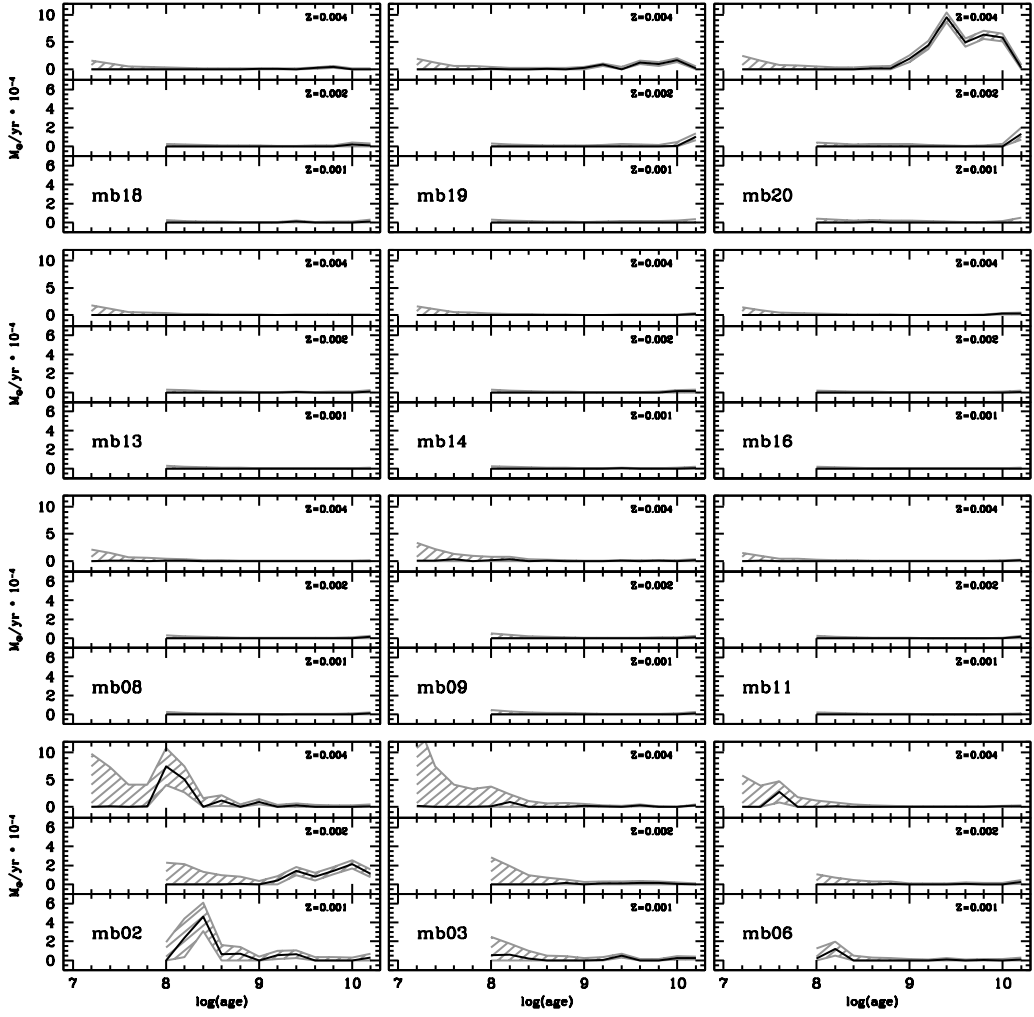}
\caption{StarFISH SFH solutions for the twelve observed fields in the 
Magellanic Bridge.  Each panel is divided into three subpanels, 
showing the star formation rate as a function of time for three 
metallicities (top to bottom: Z=0.004, Z=0.002, and Z=0.001).  
The best-fit history is plotted with a heavy line in each sub-panel, 
and the shaded gray regions indicate the estimated uncertainty of 
the solution.  Fields are presented in order of increasing right 
ascension, starting with mb02 (the field nearest the SMC) in the 
lower left, up to mb20 (the field nearest the LMC) in the upper 
right.
\label{fig:sfh} }
\end{figure}

The star formation histories of our twelve observed Magellanic Bridge 
fields are presented in Figure~\ref{fig:sfh}, and the results are 
consistent with the qualitative analysis of the CMDs presented in 
Sections~\ref{sec:youngstars} and \ref{sec:oldstars}.  Recent star 
formation has occurred only in the fields west of mb11, and old 
stellar populations are confined to the six fields nearest the SMC 
(mb02 and mb03) and LMC (mb16--mb20).  In fields mb11, mb13 and mb14 
there is no trace of a stellar population associated with the 
Magellanic system.

\begin{figure}[h]
\plotone{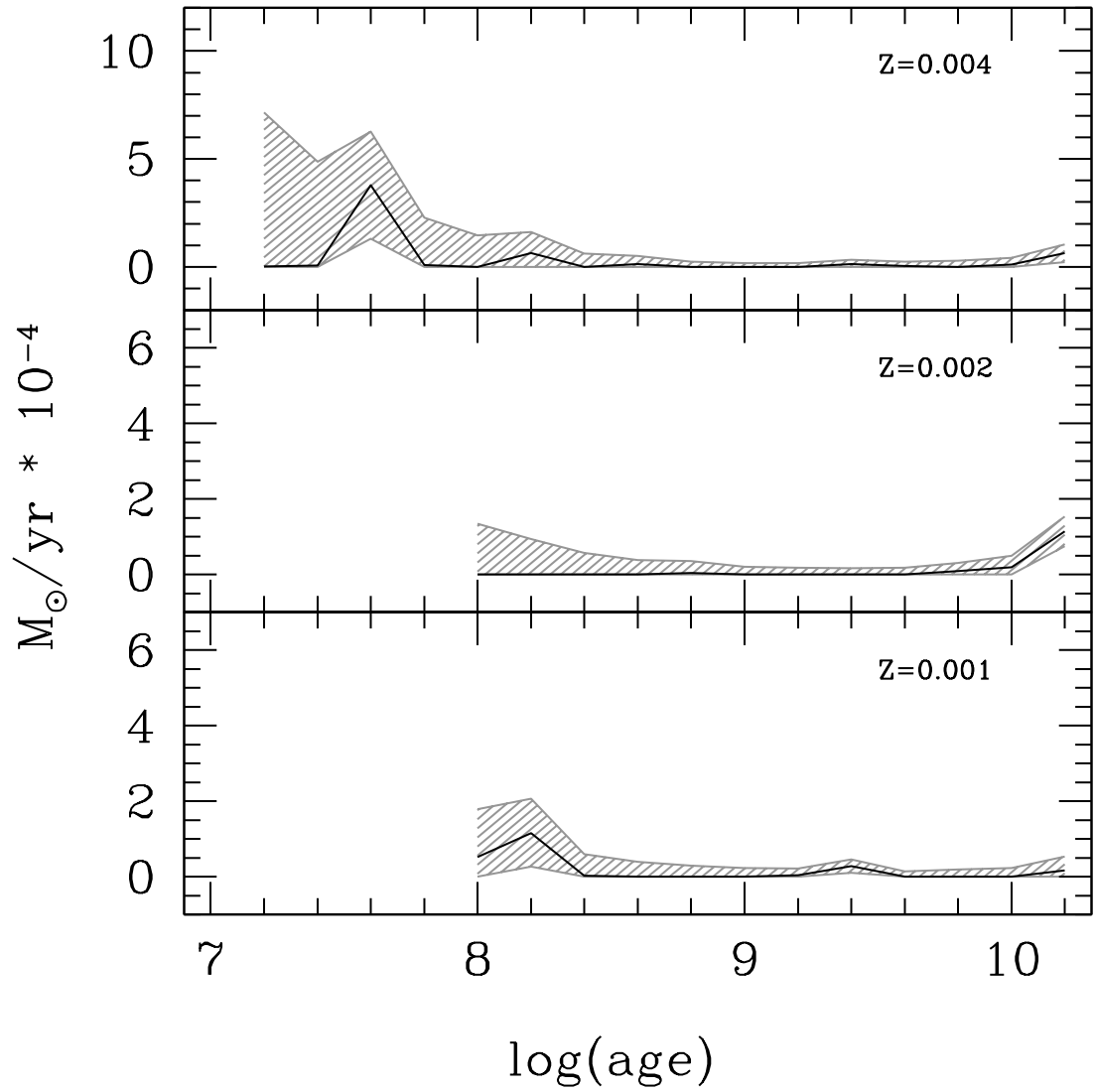}
\caption{The SFH solution for a composite population formed from 
fields mb06--mb09.  The solutions for these fields presented in 
Figure~\ref{fig:sfh} are inconclusive, due to the small number of 
stars in these fields.  By forming a composite population, we can 
better constrain the average history of these three regions.
\label{fig:comp-sfh} }
\end{figure}

Our StarFISH analysis shows that star formation in the Bridge began
around 200--300~Myr ago, and this measurement provides an important
constraint on the timing of the Bridge-forming event.  In field mb02,
the field nearest the SMC, we see a prolonged star formation episode
spanning ages 80--300~Myr, and in mb03 we see a slightly shorter
episode spanning ages 100--200~Myr.  In the other fields in which a
young stellar population is present (mb06--mb09), the star formation
rates are much lower, making ages and durations more difficult to
determine reliably.  To boost the signal, we construct a composite
population from these three fields, and determine the SFH of the
composite population (see Figure~\ref{fig:comp-sfh}).  In these more
easterly fields, we see evidence for two distinct episodes of star
formation, 160~Myr and 40~Myr ago.  It is interesting that while star
formation was active throughout the western Bridge 100--200~Myr ago,
the more recent episode 40~Myr ago was apparently confined to regions
further from the SMC.  We note that \cite{db98} also found that the
youngest inter-Cloud populations are to be found in fields eastward of
the SMC Wing.

The StarFISH solutions indicate that no significant star formation
occurred in the Bridge more recently than 40~Myr ago.  This conclusion
can be confirmed by direct inspection of the CMDs in
Figure~\ref{fig:cmds}: in no region do we see a significant number
of main sequence stars brighter than $R=15$~mag, corresponding to the
main-sequence turn off position of a 40~Myr isochrone at the distance
of the Magellanic system.

The conclusion that star formation in the Bridge largely ceased around
40~Myr seems to be at odds with a variety of previous research that
finds evidence of stellar populations much younger than this.
\cite{db98} used main-sequence isochrone fitting to conclude that the
western Bridge contains stars as young as 10--25~Myr.  \cite{mea86}
reported the discovery of DEM~171, a large circular H$\alpha$ filament
in the Bridge which is likely photoionized by one or more massive O
stars.  \cite{bs95} found that some of the clusters in their Bridge
catalog have associated emission nebulae, again implying the presence
of massive stars.  \cite{miz06} detected cold molecular clouds in the
Bridge, which demonstrates at least the potential for ongoing star
formation.

This apparent contradiction can be partly reconciled by understanding
that we are not claiming there are absolutely no stars in the Bridge
younger than 40~Myr; we find that the star-formation rate dropped off
around 40~Myr ago, and has remained consistent with zero since then.
We also note that our field selection covers the HI ridgeline of the
Magellanic Bridge uniformly (see Figure~\ref{fig:fields}), {\it
  except} for the segment around $RA=2^h$, where much of the evidence
for more recent star formation is to be found.  These explanations do
not reconcile our result with the conclusions of \cite{db98}, however.
They reported the widespread presence of stars aged 10--25 Myr in a
number of fields east of $RA=2^h$.  This is based on an analysis of
their Figure~7 (lower panel), in which theoretical isochrones are
overplotted on a composite CMD from four of their observed fields.
While the observed main sequence does appear to follow the shape of
the 10~Myr isochrone, it is clearly truncated around $M_V=-3$~mag,
whereas a 10~Myr population should have a main sequence that extends
up to $M_V=-5$~mag.  A main sequence turn-off at $M_V=-3$~mag is
consistent with the 40~Myr age that we have found for the youngest
bulk population in the Bridge.

\section{Summary}\label{sec:summary}

We have observed stellar populations in twelve fields uniformly
spanning the region between the Magellanic Clouds.  Our fields were
selected to follow the ridgeline of the \ion{H}{1} gas that forms the
Magellanic Bridge, in order to look for stars that formed {\it in
  situ} in the Bridge from gas that had already been removed from one
of the Clouds, and also for stars that were stripped from either of
the Clouds by the same tidal forces that presumably stripped the gas.

We observed the previously known young stellar population in the
western half of the inter-Cloud region, most recently characterized by
\cite{db98}, and extend on previous analyses in two key ways.  First,
we determine that the eastward extent of these stars is truncated
around $\alpha=3^h$, corresponding also to the eastward extent of the
star clusters cataloged by \cite{bs95}, and to the point at which the
\ion{H}{1} surface density falls below the critical threshold for star
formation as determined by \cite{ken89}.  Second, we use the StarFISH
program to determine quantitative star formation histories of the
young inter-Cloud population, finding that star formation in the
Bridge commenced about 200--300~Myr ago, and continued over an
extended interval, until about 40~Myr ago.  

We found no evidence for a population of tidally-stripped stars in the
inter-Cloud region, and our non-detection allows us to conclude that
the material stripped from the Clouds into the Bridge was very nearly
a pure gas, with an upper limit on the mass fraction in stars of less
than $10^{-4}$ if coincidence with the gaseous Bridge is assumed, and
0.03 otherwise.  This can potentially be understood if the
pre-collision SMC had an extended envelope of gas, surrounding a more
tightly bound stellar component.  In this case, a weak tidal
interaction might unbind the gas envelope while leaving the stellar
component undisturbed.  It is known that some dwarf galaxies have
\ion{H}{1} gas extending beyond 2--3 times the radii occupied by their
stellar populations \citep{sh96}, so perhaps this scenario is
plausible.  In fact, the recent numerical simulation of the tidal
history of the SMC by cite{yn03} included such an extended gas
envelope, in order to produce a pure-gas Magellanic Stream.
Nevertheless, the inter-Cloud region in their best model contains an
abundant population of tidally-stripped stars.  In addition, Figure~4a
of \cite{bru05} shows that the \ion{H}{1} gas in the Magellanic Bridge
appears to be contiguous with the higher-density gas in the central
regions of the SMC, which are currently abundantly populated with
stars.  If the gaseous Bridge formed via the tidal extraction of this
high-density gas from the central regions of the SMC, then the
question remains: where are the stars in the Bridge that should have
felt these same tidal forces?

Future observations may be able to address this question.  While some
kinematic measurements exist for a handful of stars in the inter-Cloud
region \citep{mau87, kid97}, radial-velocity kinematics of a truly
representative sample of the young inter-Cloud population would help
us to better understand the dynamical evolution of the Magellanic
Bridge.  A much deeper understanding of the SMC's complex
three-dimensional structure and kinematics would certainly help as
well.  These measurements (along with our current understanding of the
orbital motions of the Clouds and the Milky Way) could then be used to
motivate new detailed numerical simulations specifically targeting the
formation of the Magellanic Bridge as a pure-gas feature.  We may then
better understand how tidal features are formed during minor
harassment interactions, and what role such interactions play in
driving the evolution of the participant galaxies.

\acknowledgments 
I am very grateful for extended discussions
with Edward Olszewski on the interpretation of the data presented
here, and I would also like to thank Kurtis Williams, Abhijit Saha,
Knut Olsen, Tim Abbot and Armin Rest for their assistance with the
reduction of the Mosaic-II images.  Finally, I would like to
gratefully acknowledge the constructive comments made by the anonymous
referee.  These comments prompted the 2MASS analysis, and
substantially improved the paper.  JH is supported by NASA through
Hubble Fellowship grant HF-01160.01-A awarded by the Space Telescope
Science Institute, which is operated by the Association of
Universities for Research in Astronomy, Inc., under NASA contract NAS
5-26555.


\clearpage

\input{tab1.tex}
\input{tab2.tex}
\input{tab3.tex}
\input{tab4.tex}

\clearpage


\end{document}

%% file: tab1.tex
\begin{deluxetable}{ccccrcc}
\tabletypesize{\scriptsize}
\tablecolumns{7}
\tablewidth{0pt}

\tablecaption{Mosaic-II Fields and Exposures \label{tab:exposures}}
\tablehead{
    \colhead{Field~ID} & \colhead{Right~ascension} & \colhead{Declination} & 
        \colhead{Filter} & \colhead{$t_{exp}$} & \colhead{Observing~time} & \colhead{Airmass} \\
    \colhead{} & \colhead{} & \colhead{} &
        \colhead{} & \colhead{[sec]} & \colhead{[UT]} & \colhead{}
}

\startdata
Offset  & 00$^h$~13$^m$   & $-$79$^\circ$~59$^\prime$ & $C$ & 600 & 2006-01-04~01:19 & 1.68 \\
\nodata & \nodata & \nodata  & $R$ & 300 & 2006-01-04~01:42 & 1.72 \\
\nodata & \nodata & \nodata  & $I$ & 300 & 2006-01-04~01:56 & 1.74 \\
\nodata & \nodata & \nodata  & $C$ &  20 & 2006-01-05~01:09 & 1.67 \\
\nodata & \nodata & \nodata  & $R$ &  10 & 2006-01-05~01:11 & 1.68 \\
\nodata & \nodata & \nodata  & $I$ &  10 & 2006-01-05~01:13 & 1.68 \\
\hline
mb02    & 01$^h$~48$^m$   & $-$74$^\circ$~30$^\prime$ & $C$ & 600 & 2006-01-04~02:11 & 1.50 \\
\nodata & \nodata & \nodata  & $R$ & 300 & 2006-01-04~02:34 & 1.54 \\
\nodata & \nodata & \nodata  & $I$ & 300 & 2006-01-04~02:48 & 1.56 \\
\nodata & \nodata & \nodata  & $C$ &  20 & 2006-01-04~03:02 & 1.59 \\
\nodata & \nodata & \nodata  & $R$ &  10 & 2006-01-04~03:04 & 1.59 \\
\nodata & \nodata & \nodata  & $I$ &  10 & 2006-01-04~03:07 & 1.60 \\
\hline
mb03    & 02$^h$~00$^m$   & $-$73$^\circ$~00$^\prime$ & $C$ & 600 & 2006-01-05~01:16 & 1.40 \\
\nodata & \nodata & \nodata  & $R$ & 300 & 2006-01-05~01:42 & 1.42 \\
\nodata & \nodata & \nodata  & $I$ & 300 & 2006-01-05~01:57 & 1.44 \\
\nodata & \nodata & \nodata  & $C$ &  20 & 2006-01-05~01:40 & 1.42 \\
\nodata & \nodata & \nodata  & $R$ &  10 & 2006-01-05~01:55 & 1.44 \\
\nodata & \nodata & \nodata  & $I$ &  10 & 2006-01-05~02:10 & 1.46 \\
\hline
mb06    & 02$^h$~24$^m$   & $-$73$^\circ$~54$^\prime$ & $C$ & 600 & 2006-01-04~03:47 & 1.60 \\
\nodata & \nodata & \nodata  & $R$ & 300 & 2006-01-04~04:13 & 1.66 \\
\nodata & \nodata & \nodata  & $I$ & 300 & 2006-01-04~04:28 & 1.70 \\
\nodata & \nodata & \nodata  & $C$ &  20 & 2006-01-04~04:11 & 1.65 \\
\nodata & \nodata & \nodata  & $R$ &  10 & 2006-01-04~04:26 & 1.69 \\
\nodata & \nodata & \nodata  & $I$ &  10 & 2006-01-04~04:42 & 1.73 \\
\hline
mb08    & 02$^h$~42$^m$   & $-$73$^\circ$~30$^\prime$ & $C$ & 600 & 2006-01-05~02:13 & 1.42 \\
\nodata & \nodata & \nodata  & $R$ & 300 & 2006-01-05~02:38 & 1.45 \\
\nodata & \nodata & \nodata  & $I$ & 300 & 2006-01-05~02:53 & 1.47 \\
\nodata & \nodata & \nodata  & $C$ &  20 & 2006-01-05~02:36 & 1.45 \\
\nodata & \nodata & \nodata  & $R$ &  10 & 2006-01-05~02:51 & 1.47 \\
\nodata & \nodata & \nodata  & $I$ &  10 & 2006-01-05~03:07 & 1.49 \\
\hline
mb09    & 03$^h$~00$^m$   & $-$73$^\circ$~30$^\prime$ & $C$ & 600 & 2006-01-04~04:43 & 1.64 \\
\nodata & \nodata & \nodata  & $R$ & 300 & 2006-01-04~05:09 & 1.71 \\
\nodata & \nodata & \nodata  & $I$ & 300 & 2006-01-04~05:24 & 1.75 \\
\nodata & \nodata & \nodata  & $C$ &  20 & 2006-01-04~05:07 & 1.70 \\
\nodata & \nodata & \nodata  & $R$ &  10 & 2006-01-04~05:22 & 1.74 \\
\nodata & \nodata & \nodata  & $I$ &  10 & 2006-01-04~05:38 & 1.79 \\
\hline
mb11    & 03$^h$~18$^m$   & $-$74$^\circ$~00$^\prime$ & $C$ & 600 & 2006-01-05~03:11 & 1.46 \\
\nodata & \nodata & \nodata  & $R$ & 300 & 2006-01-05~03:37 & 1.49 \\
\nodata & \nodata & \nodata  & $I$ & 300 & 2006-01-05~03:52 & 1.51 \\
\nodata & \nodata & \nodata  & $C$ &  20 & 2006-01-05~03:35 & 1.49 \\
\nodata & \nodata & \nodata  & $R$ &  10 & 2006-01-05~03:50 & 1.51 \\
\nodata & \nodata & \nodata  & $I$ &  10 & 2006-01-05~04:06 & 1.54 \\
\hline
mb13    & 03$^h$~36$^m$   & $-$74$^\circ$~30$^\prime$ & $C$ & 600 & 2006-01-04~05:52 & 1.74 \\
\nodata & \nodata & \nodata  & $R$ & 300 & 2006-01-04~06:17 & 1.81 \\
\nodata & \nodata & \nodata  & $I$ & 300 & 2006-01-04~06:33 & 1.86 \\
\nodata & \nodata & \nodata  & $C$ &  20 & 2006-01-04~06:15 & 1.80 \\
\nodata & \nodata & \nodata  & $R$ &  10 & 2006-01-04~06:31 & 1.85 \\
\nodata & \nodata & \nodata  & $I$ &  10 & 2006-01-04~06:46 & 1.91 \\
\hline
mb14    & 03$^h$~42$^m$   & $-$73$^\circ$~18$^\prime$ & $C$ & 600 & 2006-01-05~04:20 & 1.51 \\
\nodata & \nodata & \nodata  & $R$ & 300 & 2006-01-05~04:45 & 1.55 \\
\nodata & \nodata & \nodata  & $I$ & 300 & 2006-01-05~05:00 & 1.59 \\
\nodata & \nodata & \nodata  & $C$ &  20 & 2006-01-05~04:43 & 1.55 \\
\nodata & \nodata & \nodata  & $R$ &  10 & 2006-01-05~04:58 & 1.58 \\
\nodata & \nodata & \nodata  & $I$ &  10 & 2006-01-05~05:14 & 1.62 \\
\hline
mb16    & 03$^h$~54$^m$   & $-$75$^\circ$~00$^\prime$ & $C$ & 600 & 2006-01-05~05:17 & 1.62 \\
\nodata & \nodata & \nodata  & $R$ & 300 & 2006-01-05~05:42 & 1.68 \\
\nodata & \nodata & \nodata  & $I$ & 300 & 2006-01-05~05:57 & 1.73 \\
\nodata & \nodata & \nodata  & $C$ &  20 & 2006-01-05~05:40 & 1.68 \\
\nodata & \nodata & \nodata  & $R$ &  10 & 2006-01-05~05:56 & 1.71 \\
\nodata & \nodata & \nodata  & $I$ &  10 & 2006-01-05~06:18 & 1.77 \\
\hline
mb18    & 04$^h$~12$^m$   & $-$75$^\circ$~00$^\prime$ & $C$ & 600 & 2006-01-04~06:48 & 1.80 \\
\nodata & \nodata & \nodata  & $R$ & 300 & 2006-01-04~07:14 & 1.88 \\
\nodata & \nodata & \nodata  & $I$ & 300 & 2006-01-04~07:29 & 1.93 \\
\nodata & \nodata & \nodata  & $C$ &  20 & 2006-01-04~07:12 & 1.87 \\
\nodata & \nodata & \nodata  & $R$ &  10 & 2006-01-04~07:27 & 1.93 \\
\nodata & \nodata & \nodata  & $I$ &  10 & 2006-01-04~07:42 & 1.98 \\
\hline
mb19    & 04$^h$~30$^m$   & $-$75$^\circ$~00$^\prime$ & $C$ & 600 & 2006-01-05~06:32 & 1.72 \\
\nodata & \nodata & \nodata  & $R$ & 300 & 2006-01-05~06:57 & 1.78 \\
\nodata & \nodata & \nodata  & $I$ & 300 & 2006-01-05~07:13 & 1.83 \\
\nodata & \nodata & \nodata  & $C$ &  20 & 2006-01-05~06:55 & 1.78 \\
\nodata & \nodata & \nodata  & $R$ &  10 & 2006-01-05~07:11 & 1.82 \\
\nodata & \nodata & \nodata  & $I$ &  10 & 2006-01-05~07:26 & 1.88 \\
\hline
mb20    & 04$^h$~48$^m$   & $-$74$^\circ$~30$^\prime$ & $C$ & 600 & 2006-01-05~07:28 & 1.82 \\
\nodata & \nodata & \nodata  & $R$ & 300 & 2006-01-05~07:54 & 1.91 \\
\nodata & \nodata & \nodata  & $I$ & 300 & 2006-01-05~08:09 & 1.97 \\
\nodata & \nodata & \nodata  & $C$ &  20 & 2006-01-05~07:52 & 1.90 \\
\nodata & \nodata & \nodata  & $R$ &  10 & 2006-01-05~08:07 & 1.96 \\
\nodata & \nodata & \nodata  & $I$ &  10 & 2006-01-05~08:22 & 2.02 \\
\enddata
\end{deluxetable}

%% file: tab2.tex
\begin{deluxetable}{crrcrccr}
\tabletypesize{\scriptsize}
\tablecolumns{8}
\tablewidth{0pt}

\tablecaption{Standard Field Observations \label{tab:standards}}
\tablehead{
    \colhead{Field~ID} & \colhead{Right~ascension} & \colhead{Declination} & 
        \colhead{Filter} & \colhead{$t_{exp}$} & \colhead{Observing~time} & 
        \colhead{Airmass} & \colhead{$N_{standards}$} \\
    \colhead{} & \colhead{} & \colhead{} &
        \colhead{} & \colhead{[sec]} & \colhead{} & \colhead{} & \colhead{}
}

\startdata
SA~92    & 00$^h$~55$^m$ &    00$^\circ$~40$^\prime$ & $C$ &  20 & 2006-01-04~00:53 & 1.38 &   5 \\
\nodata  & \nodata & \nodata  & $R$ &  10 & 2006-01-04~00:55 & 1.38 &  89 \\
\nodata  & \nodata & \nodata  & $I$ &  10 & 2006-01-04~00:57 & 1.39 & 132 \\
SA~92  & 00$^h$~55$^m$ &    00$^\circ$~40$^\prime$ & $C$ &  20 & 2006-01-04~03:13 & 3.01 &   6 \\
\nodata  & \nodata & \nodata  & $R$ &  10 & 2006-01-04~03:15 & 3.08 &  98 \\
\nodata  & \nodata & \nodata  & $I$ &  10 & 2006-01-04~03:17 & 3.14 & 147 \\
SA~101   & 09$^h$~57$^m$ & $-$00$^\circ$~20$^\prime$ & $C$ &  20 & 2006-01-04~05:43 & 1.34 &   6 \\
\nodata  & \nodata & \nodata  & $R$ &  10 & 2006-01-04~05:45 & 1.33 &  51 \\
\nodata  & \nodata & \nodata  & $I$ &  10 & 2006-01-04~05:47 & 1.33 &  79 \\
SA~101   & 09$^h$~57$^m$ & $-$00$^\circ$~20$^\prime$ & $C$ &  20 & 2006-01-04~07:47 & 1.15 &   6 \\
\nodata  & \nodata & \nodata & $R$ &  10 & 2006-01-04~07:50 & 1.15 &  49 \\
\nodata  & \nodata & \nodata & $I$ &  10 & 2006-01-04~07:51 & 1.15 &  82 \\
SA~98    & 06$^h$~52$^m$ & $-$00$^\circ$~24$^\prime$ & $C$ &  20 & 2006-01-04~07:55 & 1.73 &  15 \\
\nodata  & \nodata & \nodata & $R$ &  10 & 2006-01-04~07:57 & 1.75 & 662 \\
\nodata  & \nodata & \nodata & $I$ &  10 & 2006-01-04~07:59 & 1.77 & 888 \\
NGC~2298 & 06$^h$~49$^m$ & $-$36$^\circ$~00$^\prime$ & $C$ &  20 & 2006-01-04~08:02 & 1.36 & \nodata \\
\nodata  & \nodata & \nodata & $R$ &  10 & 2006-01-04~08:04 & 1.37 & 515 \\
\nodata  & \nodata & \nodata & $I$ &  10 & 2006-01-04~08:06 & 1.38 &  42 \\
\hline
SA~92    & 00$^h$~55$^m$ &    00$^\circ$~40$^\prime$ & $C$ &  20 & 2006-01-05~01:01 & 1.42 &   7 \\
\nodata  & \nodata & \nodata & $R$ &  10 & 2006-01-05~01:03 & 1.43 &  97 \\
\nodata  & \nodata & \nodata & $I$ &  10 & 2006-01-05~01:05 & 1.44 & 144 \\
SA~101   & 09$^h$~57$^m$ & $-$00$^\circ$~20$^\prime$ & $C$ &  20 & 2006-01-05~04:11 & 1.91 &   6 \\
\nodata  & \nodata & \nodata & $R$ &  10 & 2006-01-05~04:13 & 1.88 &  55 \\
\nodata  & \nodata & \nodata & $I$ &  10 & 2006-01-05~04:15 & 1.86 &  80 \\
SA~101   & 09$^h$~57$^m$ & $-$00$^\circ$~20$^\prime$ & $C$ &  20 & 2006-01-05~06:23 & 1.22 &   7 \\
\nodata  & \nodata & \nodata & $R$ &  10 & 2006-01-05~06:25 & 1.22 &  52 \\
\nodata  & \nodata & \nodata & $I$ &  10 & 2006-01-05~06:27 & 1.22 &  77 \\
\enddata
\end{deluxetable}

%% file: tab3.tex
\begin{deluxetable}{cccccccc}
\tabletypesize{\scriptsize}
\tablecolumns{8}
\tablewidth{0pt}

\tablecaption{Photometric Calibration Parameters \label{tab:photcalib}}
\tablehead{
    \colhead{Filter} & \colhead{CCD} & \colhead{Extincton} & 
        \colhead{Ext. Unc.} & \colhead{ZP} & \colhead{ZP Unc.} 
        & \colhead{Color~Term} & \colhead{ Col. Term Unc.} \\
}

\startdata
\multicolumn{8}{c}{Night 1} \\
\hline
 $C$ & 1--8\tnm{a} &  0.275 &  0.0015 &   0.074 &  0.0200 &  -0.100 &  0.0100 \\
 $R$ & 1\tnm{b} &  0.072 &  0.0010 &  -0.690 &  0.0200 &  -0.032 &  0.0130 \\
 $R$ & 2 &  0.072 &  0.0010 &  -0.706 &  0.0045 &  -0.024 &  0.0077 \\
 $R$ & 3 &  0.072 &  0.0010 &  -0.684 &  0.0023 &  -0.042 &  0.0039 \\
 $R$ & 4 &  0.072 &  0.0010 &  -0.670 &  0.0045 &  -0.049 &  0.0083 \\
 $R$ & 5\tnm{b} &  0.072 &  0.0010 &  -0.682 &  0.0200 &  -0.035 &  0.0130 \\
 $R$ & 6 &  0.072 &  0.0010 &  -0.680 &  0.0033 &  -0.044 &  0.0050 \\
 $R$ & 7 &  0.072 &  0.0010 &  -0.666 &  0.0027 &  -0.052 &  0.0042 \\
 $R$ & 8 &  0.072 &  0.0010 &  -0.703 &  0.0027 &  -0.030 &  0.0050 \\
 $I$ & 1 &  0.040 &  0.0009 &  -0.029 &  0.0185 &  -0.003 &  0.0140 \\
 $I$ & 2 &  0.040 &  0.0009 &   0.009 &  0.0022 &   0.007 &  0.0016 \\
 $I$ & 3 &  0.040 &  0.0009 &   0.008 &  0.0020 &  -0.001 &  0.0016 \\
 $I$ & 4 &  0.040 &  0.0009 &   0.014 &  0.0038 &   0.005 &  0.0041 \\
 $I$ & 5\tnm{b} &  0.040 &  0.0009 &  -0.006 &  0.0230 &   0.003 &  0.0090 \\
 $I$ & 6 &  0.040 &  0.0009 &  -0.011 &  0.0020 &   0.012 &  0.0015 \\
 $I$ & 7 &  0.040 &  0.0009 &   0.010 &  0.0015 &   0.003 &  0.0013 \\
 $I$ & 8 &  0.040 &  0.0009 &  -0.003 &  0.0031 &   0.010 &  0.0022 \\
\hline
\multicolumn{8}{c}{Night 2} \\
\hline
 $C$ & 1--8\tnm{a} &  0.299 &  0.0012 &   0.026 &  0.0200 &  -0.087 &  0.0100 \\
 $R$ & 1\tnm{b} &  0.090 &  0.0010 &  -0.731 &  0.0200 &  -0.021 &  0.0130 \\
 $R$ & 2 &  0.090 &  0.0010 &  -0.732 &  0.0078 &  -0.022 &  0.0140 \\
 $R$ & 3 &  0.090 &  0.0010 &  -0.707 &  0.0030 &  -0.032 &  0.0050 \\
 $R$ & 4 &  0.090 &  0.0010 &  -0.727 &  0.0088 &  -0.047 &  0.0176 \\
 $R$ & 5\tnm{b} &  0.090 &  0.0010 &  -0.723 &  0.0200 &  -0.024 &  0.0130 \\
 $R$ & 6 &  0.090 &  0.0010 &  -0.710 &  0.0055 &  -0.030 &  0.0082 \\
 $R$ & 7 &  0.090 &  0.0010 &  -0.727 &  0.0067 &  -0.022 &  0.0094 \\
 $R$ & 8 &  0.090 &  0.0010 &  -0.748 &  0.0043 &  -0.020 &  0.0079 \\
 $I$ & 1\tnm{b} &  0.061 &  0.0007 &  -0.038 &  0.0220 &   0.007 &  0.0130 \\
 $I$ & 2 &  0.061 &  0.0007 &  -0.011 &  0.0025 &   0.002 &  0.0017 \\
 $I$ & 3 &  0.061 &  0.0007 &  -0.030 &  0.0047 &   0.004 &  0.0038 \\
 $I$ & 4 &  0.061 &  0.0007 &  -0.003 &  0.0111 &  -0.019 &  0.0109 \\
 $I$ & 5\tnm{b} &  0.061 &  0.0007 &  -0.033 &  0.0200 &   0.001 &  0.0130 \\
 $I$ & 6 &  0.061 &  0.0007 &  -0.029 &  0.0026 &   0.018 &  0.0021 \\
 $I$ & 7 &  0.061 &  0.0007 &  -0.021 &  0.0031 &  -0.002 &  0.0021 \\
 $I$ & 8 &  0.061 &  0.0007 &  -0.029 &  0.0028 &   0.011 &  0.0020 \\

\enddata

\tablenotetext{a}{There were to few $C$ standards to support an
  independent transformation solution for each CCD.}
\tablenotetext{b}{There were too few standards in this CCD to support
  an independent transformation solution.  The zeropoint and color
  term values for this CCD are bootstrapped from the values published
  at the CTIO website.}
\end{deluxetable}

%% file: tab4.tex
\begin{deluxetable}{ccccccccc}
\tabletypesize{\scriptsize}
\tablecolumns{9}
\tablewidth{0pt}
 
\tablecaption{Photometry of Stars in Magellanic Bridge Fields \label{tab:catalog}}
\tablehead{
    \colhead{Object~ID} & 
    \colhead{Right~ascension} & \colhead{Declination} & 
    \colhead{$C$} & \colhead{$\sigma_C$} & 
    \colhead{$R$} & \colhead{$\sigma_R$} & 
    \colhead{$I$} & \colhead{$\sigma_I$} \\
    \colhead{} & \colhead{} & \colhead{} &
    \colhead{[mag]} & \colhead{[mag]} & \colhead{[mag]} & \colhead{[mag]} &
    \colhead{[mag]} & \colhead{[mag]} 
}
 
\startdata
\multicolumn{9}{c}{Region mb02} \\
\cline{1-9}
   1 & 01$^h$ 43$^m$ 10.7$^s$ & -74$^\circ$ 37$^\prime$ 51$^{\prime\prime}$ &  19.187 &   0.037 &  19.416 &   0.028 &  19.385 &   0.039 \\ 
   2 & 01$^h$ 43$^m$ 10.7$^s$ & -74$^\circ$ 38$^\prime$ 52$^{\prime\prime}$ & \nodata & \nodata &  21.070 &   0.087 &  20.796 &   0.115 \\ 
   3 & 01$^h$ 43$^m$ 10.9$^s$ & -74$^\circ$ 37$^\prime$ 36$^{\prime\prime}$ &  18.075 &   0.037 &  18.486 &   0.026 &  18.482 &   0.032 \\ 
   4 & 01$^h$ 43$^m$ 11.0$^s$ & -74$^\circ$ 44$^\prime$ 39$^{\prime\prime}$ &  21.457 &   0.116 &  20.501 &   0.052 & \nodata & \nodata \\ 
   5 & 01$^h$ 43$^m$ 11.2$^s$ & -74$^\circ$ 33$^\prime$ 20$^{\prime\prime}$ & \nodata & \nodata &  20.255 &   0.045 &  19.720 &   0.059 \\ 
   6 & 01$^h$ 43$^m$ 11.2$^s$ & -74$^\circ$ 36$^\prime$ 26$^{\prime\prime}$ &  20.416 &   0.060 &  18.901 &   0.023 &  18.379 &   0.033 \\ 
   7 & 01$^h$ 43$^m$ 11.2$^s$ & -74$^\circ$ 37$^\prime$ 22$^{\prime\prime}$ &  20.728 &   0.065 &  19.807 &   0.041 &  19.424 &   0.038 \\ 
   8 & 01$^h$ 43$^m$ 11.5$^s$ & -74$^\circ$ 36$^\prime$ 02$^{\prime\prime}$ & \nodata & \nodata &  19.034 &   0.030 &  18.538 &   0.033 \\ 
   9 & 01$^h$ 43$^m$ 11.8$^s$ & -74$^\circ$ 48$^\prime$ 51$^{\prime\prime}$ &  21.007 &   0.071 &  20.693 &   0.078 & \nodata & \nodata \\ 
  10 & 01$^h$ 43$^m$ 11.9$^s$ & -74$^\circ$ 45$^\prime$ 31$^{\prime\prime}$ &  17.561 &   0.035 &  15.270 &   0.032 & \nodata & \nodata \\ 
\enddata
\tablecomments{The complete version of this table is in the
electronic edition of the Journal.  The printed edition contains
only a sample.}
\end{deluxetable}